\documentclass{jetpl}
\pdfoutput=1
\twocolumn
\lat

\renewcommand\vec{\mathbf}           % denote vectors with bold
\newcommand{\uGrad}{\vec{\nabla}}    % gradient operator

\newcommand{\uiiint}{\int\!\!\!\int\!\!\!\int}
\newcommand{\ud}{\,{\mathrm{d}}}
\newcommand{\uRe}{\Real}
\newcommand{\uIm}{\Imag}

\newcommand{\uvHD}{\vec{H}_\mathrm{D}} % demag. field
\newcommand{\uvhD}{\vec{h}_\mathrm{D}}% reduced demag. field
\newcommand{\uLe}{L_\mathrm{E}}       % exchange length
\newcommand{\uvM}{\vec{M}}           % magnetization vector
\newcommand{\uvm}{\vec{m}}           % reduced magnetization vector
\newcommand{\umi}{m_i}               % reduced magnetization component, i=X,Y,Z
\newcommand{\umx}{m_{\mathrm{X}}}      % reduced magnetization component X
\newcommand{\umy}{m_{\mathrm{Y}}}      % reduced magnetization component Y
\newcommand{\umz}{m_{\mathrm{Z}}}      % reduced magnetization component Z
\newcommand{\uMs}{M_{\mathrm{S}}}      % saturation magnetization
\newcommand{\uMssq}{M_{\mathrm{S}}^2}  % saturation magnetization squared
\newcommand{\umuZ}{\mu_0}            % permeability of vacuum
\newcommand{\uvr}{\vec{r}}           % point in space
\newcommand{\uX}{X}                  % Cartesian X coordinate
\newcommand{\uY}{Y}                  % Cartesian Y coordinate
\newcommand{\uZ}{Z}                  % Cartesian Z coordinate

\newcommand{\uf}{f}                  % complex function f
\newcommand{\ufc}{\overline{f}}      % -"- conjugated
\newcommand{\uw}{w}                  % complex function w
\newcommand{\uwc}{\overline{w}}      % -"- conjugated
\newcommand{\uz}{z}                  % complex variable
\newcommand{\uzc}{\overline{z}}      % -"- conjugated
\usepackage{tikz}
\usetikzlibrary{tikzmark,calc}

\title{Topological memory with multiply-connected planar magnetic nanoelements}

\rtitle{Topological memory with multiply-connected planar magnetic
nanoelements}

\sodtitle{Topological memory with multiply-connected planar magnetic nanoelements}

\author{K.\,L.\,Metlov\/\thanks{e-mail: metlov@donfti.ru}}

\rauthor{METLOV}

\sodauthor{Metlov}

\address{Donetsk Institute for Physics and Engineering, 72 R.~Luxembourg str., 283050~Donetsk, DPR, Russian Federation\\~\\
Institute for Numerical Mathematics RAS, 8~Gubkina str., 119991~Moscow GSP-1, Russian Federation}

\dates{11 April 2023}{6 June 2023}

\abstract{A coding scheme is introduced, allowing to store a set of linked bit strings in planar magnetic nanoelements with holes. Analytical expressions for the corresponding magnetization distributions are developed up to a homotopy and the specific examples are given for doubly- and triply-connected cases. The energy barriers, protecting the information-bearing states, are discussed. Compared to a set of disparate simply-connected nanoelements of the same total connectivity, the nanoelements with holes can hold much more information due to the possibility of linking the individual bits.}

\PACS{75.60.Ch, 75.70.Kw, 85.70.Kh}

\begin{document}
\maketitle

Magnetic memory is one of the staple applications of modern magnetism. A recent concept of the ``racetrack'' memory~\cite{ParkinRacetrack2008} is currently guiding the development of skyrmionics~\cite{FCS2013}, based on ferromagnetic~\cite{TMZTCF2014} and, more recently, on antiferromagnetic skyrmions~\cite{QLCWF2022}. In ``racetrack'' the information is coded as a sequence of magnetic domain walls/skyrmions in a long nano-stripe. Here this concept is generalized to multiply connected planar nanoelements. While it becomes impossible to move bits sequentially, the state of the element can still be assessed via the resonant frequency response, like it was demonstrated in~\cite{JNFPTSB2012}.

The coding is topological, in a sense that the stored information is robust with respect to continuous transformations (subject to the boundary conditions, of course) of the corresponding magnetization distribution. Skyrmion ``racetrack'' also has a similar topological protection, but the present coding scheme allows (for connectivity $>1$) storing much more information.

It is well known that magnetization states in an infinite 2D ferromagnet (or sufficiently thin film) can be subdivided into topological classes~\cite{BP75}. The distributions within the same class are equivalent up to a continuous deformation (homotopy), while converting between the distributions of different classes requires creating a magnetization vector field singularity. In the continuum model of a ferromagnet, such a singularity has an infinite energy and,  thus, conversion between distributions of different classes is impossible. The classes can be numbered by integers and the corresponding magnetization distributions expressed via rational functions of a complex variable~\cite{BP75}. The topological index (number of the class) is the total count of vortex-antivortex pairs in the magnetization vector field or the skyrmion number. Although in finite nanoelements, it is possible to also have incomplete vortex-antivortex pairs (an extra vortex~\cite{WWBPMW02} or  antivortex~\cite{MEGKRGUFMCSP2010}) inside the element and thus the skyrmion number may assume half-integer values.

While the topological protection of the index makes the corresponding states robust, there is a limitation that the integer (or half-integer) index's algebra is abelian. It only matters -- how many vortices/antivortices are inside the element. Or, if the maximum number of skyrmions the element can hold is $n$, it can only store around $\log_2 n$ bits of information. A common approach to store more bits is to make the magnetic medium multiply-connected by employing many (let's say $N$) separate elements, which may contain (or not) vortices/skyrmions independently. The algebra of topological index becomes non-abelian, since, in addition to their total number, location of skyrmions becomes important. This allows to store around $(1+n)^N$ different states. An alternative, explored below, is to have a single planar nano-element with holes. Such an element is multiply-connected too and its topological index can also be made non-abelian, but, in addition to storing multiple bit strings, it may also accommodate links between the individual bits of the strings, which further expands the set of different topologically protected states.

Our starting point is a continual Hamiltonian of a ferromagnet with the normalized by $\umuZ\uMssq$ energy density
\begin{equation}
 \label{eq:energy}
  e=\uLe^2\sum\limits_{i=\uX,\uY,\uZ}|\uGrad\umi(\uvr)|^2 + \uvhD(\uvr,\{\uvm(\uvr)\})\cdot\uvm(\uvr),
\end{equation}
where $\uLe=\sqrt{C/(\umuZ \uMssq)}$, $C=2A$ is the exchange stiffness, $\umuZ$ is the permeability of vacuum, $\uMs$ is the saturation magnetization of the ferromagnet, $\uvm(\uvr)=\uvM(\uvr)/\uMs$ is the normalized local magnetization at a location $\uvr=\{\uX,\uY,\uZ\}$, $\uGrad=\{\partial/\partial\uX,\partial/\partial\uY,\partial/\partial\uZ\}$ is the gradient operator and $\uvhD=\uvHD/\uMs$ is the demagnetizing field, created by the magnetic poles of the magnetization distribution $\uvm(\uvr)$. The magnet has a shape of a (finite) planar thin film element (a generalized cylinder, whose base may potentially contain some holes) with the Cartesian coordinate $\uZ$ perpendicular to its faces (replicas of the base) and the plane $\uX$-$\uY$ parallel to them. The element is assumed to be thin enough that the magnetization distribution is uniform across its thickness ($\partial\uvm/\partial\uZ=0$) and, thus, identical on its top and bottom faces. The side of the nanoelement is the surface, parallel to $\uZ$ axis, running between the nanoelement faces through their boundary. Multiply-connected nanoelements (having base with holes) have number of their boundary components (boundaries, sides) equal to their connectivity.

%In the present consideration we will follow the approximate approach~\cite{M10}, which constructs the magnetization distributions, based on the exact treatment of the exchange energy functional, in a way to minimize the amount of magnetic poles. In particular, it sets the boundary condition of no normal components to the side of the planar nanoelement (eliminating the side magnetic poles completely). This minimalistic model well captures the topological properties of magnetization distributions and reproduces (quantitatively in many cases) many static and dynamic properties of magnetic vortices in nanocylinders.

Metastable magnetization distributions $\uvm(\uvr)$ extremizing the volume integral of~\eqref{eq:energy} can be expressed approximately~\cite{M10} in terms of a complex function $\uw(\uz,\uzc)$ of complex variable $\uz=\uX+\imath\uY$, $\uzc=\uX - \imath\uY$, $\imath=\sqrt{-1}$. Specifically, the reduced magnetization vector components $|\uvm|=1$ are given via a stereographic projection $\{\umx+\imath\umy,\umz\}=\{2\uw,1-\uw\uwc\}/(1+\uw\uwc)$. In case of skyrmions, the function $\uw$ is holomorphic ($\partial\uw/\partial\uzc=0$). In a more general case of magnetic vortices (merons~\cite{G78}) it is not, but can be expressed via a holomorphic function $\uf(\uz)$ in the piecewise continuous form:
\begin{equation}
\label{eq:wpieciwise}
\uw(\uz,\uzc) = 
\begin{cases} 
      \uf(\uz)/c_1 & |\uf(\uz)| \leq c_1 \\
      \sqrt{\uf(\uz)/\ufc(\uzc)} & c_1 < |\uf(\uz)| < c_2 \\
      \uf(\uz)/c_2 & |\uf(\uz)| \geq c_2 
   \end{cases},
\end{equation}
where real scalars $c_1$ and $c_2$ control the size of vortex and anti-vortex cores. Zeroes of $f(z)$ correspond to the centers of vortices (skyrmions), while its poles to the centers of antivortices (antiskyrmions). The case of pure skyrmions is realized when $c_1=c_2$. The function $f(z)$ can be obtained~\cite{M10} as a solution of the famous Riemann-Hilbert problem of finding a holomorphic function with no normal components to the boundary of the region (corresponding to the absolute minimum of the energy of the magnetic poles at the side of the particle). The solution of this (homogeneous) problem usually depends on a number of scalar parameters, related to vortex and antivortex positions. To get a specific stable magnetization distribution in an element of a particular size and made of a specific material the total energy $E=\uiiint e\ud^3r$ needs to be computed and minimized over these scalar parameters as well as $c_1$ and $c_2$. This Ritz approach is much simpler than the original variational problem.

All the metastable magnetic states, predicted by this model (in simply- and multiply- connected cases), consist of two types of magnetic vortices and antivortices -- ones that are located within the face of the nanoelement and ones, that are situated exactly on its boundary/boundaries. To introduce the coding scheme, let us assume for now that these two types of vortices/antivortices never mix and that the vortex/antivortex created inside the face, stays inside and the one, created at the boundary, stays at the boundary.

This suggests the idea to code the information by the sequence of half-vortices and half-antivortices at the boundaries of the multiply-connected element. Because making a full circle around any boundary must correspond to a full rotation (or several) of the magnetization vector, there is always an even number of half-vortices and half-antivortices at every boundary. Thus, to be able to code arbitrary strings, it is convenient to represent a symbol (let's call it $A$) by the pair of neighbouring half-vortices at the boundary and the anti-symbol ($\overline{A}$) by the pair of half-antivortices. This way each boundary (including the outer one) can contain a string of symbols and anti-symbols and the number of such strings is equal to the number of boundaries. Furthermore, because each half-vortex and half-antivortex at the boundary is a start of a domain wall, going inside the face, there are two possibilities: either the wall starts and ends at the same boundary or at a different boundary. The latter can be used to store links between the symbols, by extending a pair of domain walls between the half-vortices/antivortices, coding a symbol at one boundary, to another pair, coding a symbol on another boundary.

This way, a multiply connected element can hold a number of bit strings with links between some of the individual bits. Because of the links, the amount of stored information inside a multiply-connected element with holes surpasses the amount of information, stored within the number of separate elements of the same connectivity.

To elucidate the above scheme, let us now compute a couple of examples. The analytical function $f(z)$ is specified by the positions of the magnetic vortex centers and can be expressed via real meromorphic differentials~\cite{Bogatyrev2017} or explicitly in terms of the Schottky-Klein prime functions~\cite{BM15_eng}. There are constraints~\cite{BM17}, specific to multiply-connected elements, which relate positions of vortices and antivortices, so that not all of them are independent.

\begin{figure}
 \begin{center}
 \includegraphics[width=\columnwidth]{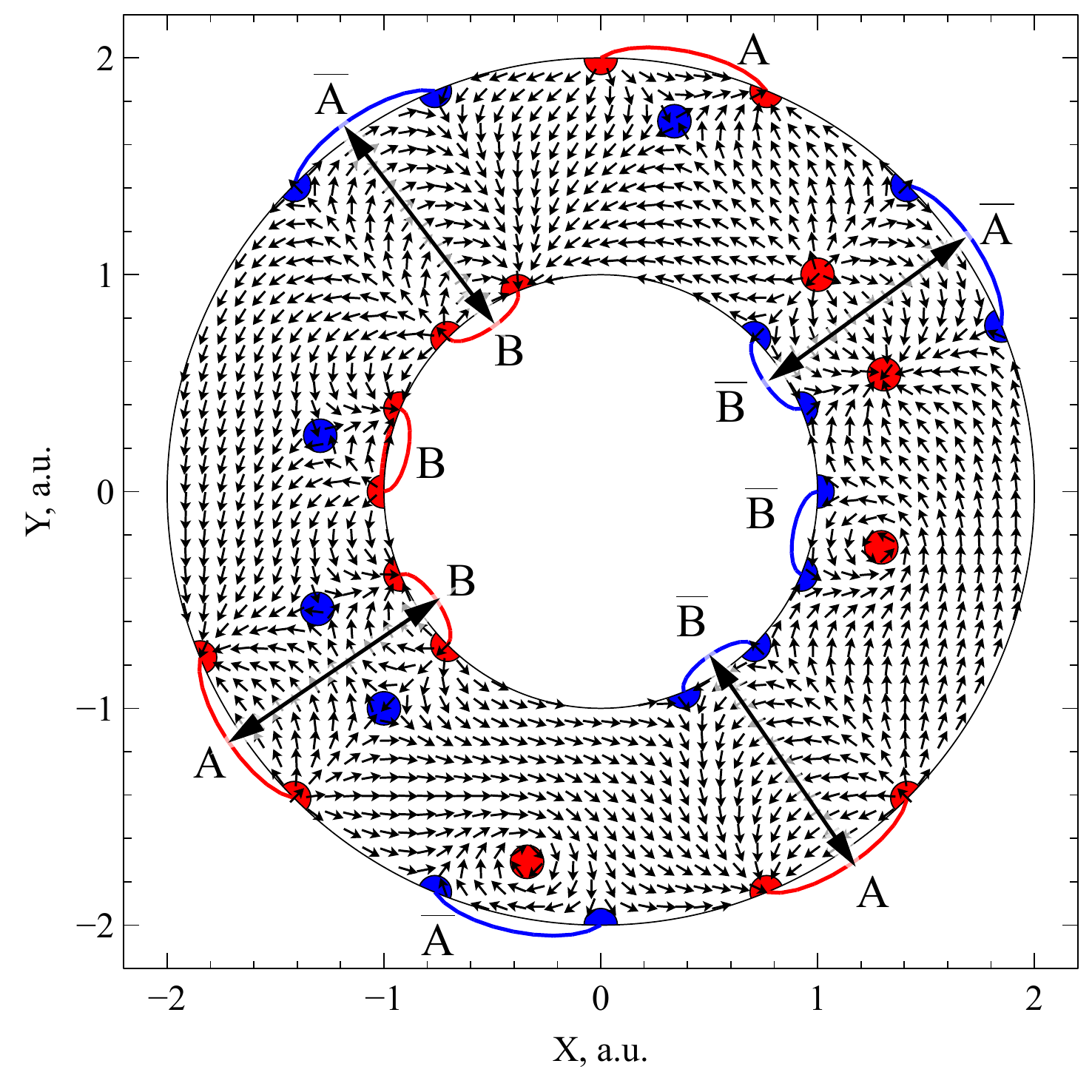}
 \vspace{-1.0cm}
 \end{center}
 \caption{\label{fig:annulus} An $R=2$ ring with $N=16$ in a configuration $f(u)=\imath F_{1,2}^{B}\allowbreak F_{3,4}^{B\overline{A}}\allowbreak F_{5,6}^{A}\allowbreak F_{7,8}^{\overline{A}\overline{B}}\allowbreak F_{9,10}^{\overline{B}}\allowbreak F_{11,12}^{\overline{B}A}\allowbreak F_{13,14}^{\overline{A}}\allowbreak F_{15,16}^{AB}$. Filled circles mark vortex and antivortex centers.}
\end{figure}

Let us start with doubly-connected planar nanoelements -- rings, and consider a planar circular concentric ring of the external radius $R$ and the internal radius $r<R$. For the sake of simplicity, we will not minimize the total magnetic energy of the particle to set the values of the parameters $c_1$ and $c_2$ and the exact locations of vortices and antivortices. They will be set arbitrarily from aesthetic considerations. For the same reason, the spatial scale will be measured in arbitrary units by setting $r=1$. Then, a ring $1\leq|z|\leq R$ can be conformally mapped onto a rectangle $0\leq\uRe u \leq1/2$, $-T/2 \leq \uIm u < T/2$ in the complex plane $u$ via $z=Z(u)=\exp(2\pi u/T)$, where $T=\pi/\log R$. The inner circle of the ring is mapped to $\uRe u=0$ and the outer to $\uRe u=1/2$. The magnetization distributions then correspond to holomorphic functions $f(u)$ (or $f(z)=f(u)Z^\prime(u)$ with $u=Z^{(-1)}(z)$ on the original $z$ complex plane), which have the period $T$ along the imaginary axis. These can be expressed in terms of elliptic theta functions~\cite{Bogatyrev2017} via positions of their zeroes (vortex centers) and poles (antivortex centers). The topological constraints will be satisfied if the sum of azimuthal angles of all vortices is equal to that of antivortices~\cite{BM17}. This is easy to ensure while laying them out with an equal angular steps along the ring: in $u$ space at $\uIm u = b_n = \imath T/2 - \imath T (n-1)/N$ with $1\leq n\leq N$ and $N$ being the total number of steps. The points $u=b_n$ correspond to the inner boundary of the ring, and $u=a_n=1/2+b_n$ to the outer boundary. It is also convenient to introduce midpoints inside of the ring $c_{n_1 n_2}=1/4+(b_{n_1}+b_{n_2})/2$ and a couple of shifted (for aesthetic reasons) midpoints $d_{n_1 n_2}=3/20+c_{n_1 n_2}$, $e_{n_1 n_2}=-1/20+c_{n_1 n_2}$. Let us denote the pair of neighbouring half-vortices on the outer boundary as $A$ and the pair of neighboring half-antivortices there as $\overline{A}$, the pairs on the inner boundary will be denoted as $B$ and $\overline{B}$ respectively. Define 8 functions:
\begin{align*}
F_{n_1 n_2}^{\overline{A}}(u)=&\frac{T_1(u-d_{n_1 n_2})T_1(u+\overline{d_{n_1 n_2}})}{T_1(u-a_{n_1})T_1(u-a_{n_2})}, \displaybreak[0]\\
F_{n_1 n_2}^{\overline{B}}(u)=&\frac{T_1(u-e_{n_1 n_2})T_1(u+\overline{e_{n_1 n_2}})}{T_1(u-b_{n_1})T_1(u-b_{n_2})}, \displaybreak[0]\\
F_{n_1 n_2}^{B\overline{A}}(u)=&-\frac{T_1(u-b_{n_1})T_1(u-\overline{b_{n_2}})}{T_1(u-a_{n_1})T_1(u-a_{n_2})}, \displaybreak[0]\\
F_{n_1 n_2}^{\overline{A}\overline{B}}(u)=& -\left(\frac{T_1(u-c_{n_1n_1})T_1(u+\overline{c_{n_1n_1}})}{T_1(u-a_{n_1})T_1(u-b_{n_1})}\right)\times \\
& \left(\frac{T_1(u-c_{n_2n_2})T_1(u+\overline{c_{n_2n_2}})}{T_1(u-a_{n_2})T_1(u-b_{n_2})}\right),
\end{align*}
$F_{n_1 n_2}^{A}=1/F_{n_1 n_2}^{\overline{A}}$, $F_{n_1 n_2}^{B} = 1/F_{n_1 n_2}^{\overline{B}}$, $F_{n_1 n_2}^{\overline{B}A} = 1/F_{n_1 n_2}^{B\overline{A}}$, $F_{n_1 n_2}^{A B} = 1/F_{n_1 n_2}^{\overline{A}\overline{B}}$,  where $T_1(u)=\theta_1(u,\imath T)$ and $\theta_1(\nu,\tau)$ is Akhiezer's theta function~\cite{akhiezer1990translation}, which can also be expressed via Jacobi's theta function as $\theta_1(\nu,\tau)=\imath e^{-\imath\pi(\nu-\tau/4)}\theta(\nu+(1-\tau)/2,\tau)$. An arbitrary configuration of bit strings on inner and outer rings with optional links between individual bits on each of the boundaries can be expressed as a product of these functions. The functions $F^{A}$,$F^{\overline{A}}$, $F^{B}$,$F^{\overline{B}}$ code the standalone bits of the strings, and the functions $F^{AB}$, $F^{\overline{A}B}$, $F^{A\overline{B}}$,$F^{\overline{A}\overline{B}}$ code the linked bits on the opposite boundaries. The example of such a configuration, showcasing all the above defined functions is shown in Fig.~\ref{fig:annulus}.
It corresponds to the following graph of symbols
\vspace{0.2cm}
\begin{equation}
 \vspace{-0.2cm}
 \label{eq:state2c}
 {\tikzmarknode{AAs}{\overline{A}}A\tikzmarknode{AAm}{\overline{A}}A\overline{A}\tikzmarknode{AAe}{A}}\qquad
 {\tikzmarknode{BBs}{B}B\tikzmarknode{BBm1}{B}\tikzmarknode{BBm2}{\overline{B}}\overline{B}\tikzmarknode{BBe}{\overline{B}}}.
\end{equation}
\tikz[remember picture, overlay]{
\draw[latex-latex,red] ([yshift=0.1em]AAs.north) to[bend left] ([yshift=0.25em]AAe.north);
\draw[latex-latex,red] ([yshift=0.1em]BBs.north) to[bend left] ([yshift=0.1em]BBe.north);
\draw[latex-latex,red] ([yshift=0.1em]AAm.north) to[bend left=12] ([yshift=0.1em]BBm2.north);
\draw[latex-latex,red] ([yshift=0.1em]AAe.east) to ([yshift=0.1em]BBs.west);
\draw[latex-latex,red] ([yshift=-0.1em]AAs.south) to[bend right=12] ([yshift=-0.1em]BBm1.south);
}
The strings are cyclic, which is denoted via additional links from their beginning to their end. This property can be removed by introducing some notches or other inhomogeneities at the boundaries, creating energy barriers for half-vortices/antivortices and marking the start of the strings. Alternatively, the mark can be set using a single link, connecting the starting bits of $A$ and $B$ strings. This makes notches unnecessary, although the possibility to use the links for information storage is then greatly reduced.

\begin{figure}
 \begin{center}
 \includegraphics[width=\columnwidth]{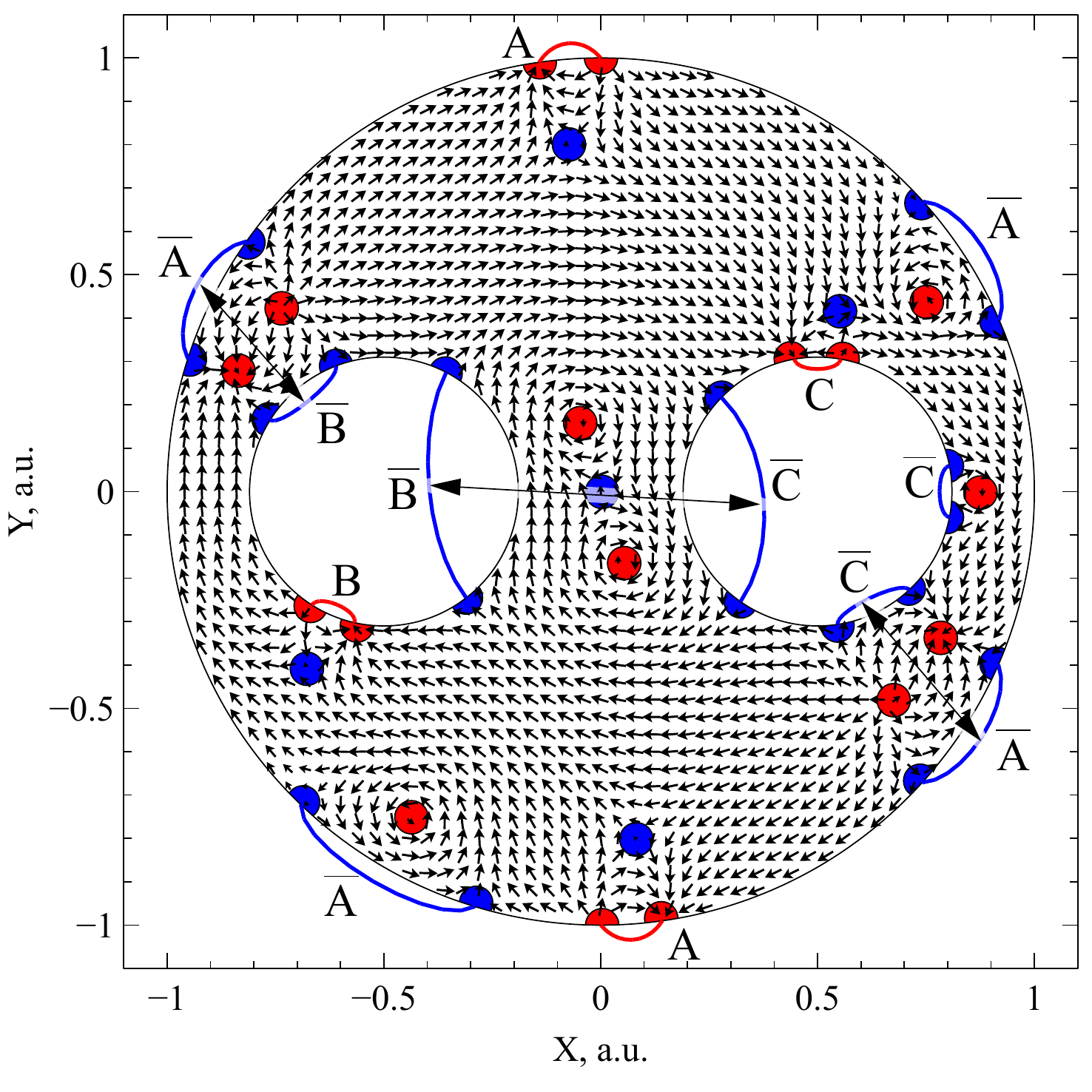}
 \vspace{-1.0cm}
 \end{center}
 \caption{\label{fig:triplyconnected} A triply connected circular planar nanoelement of the radius $R_3=1$ with the circular holes of equal radii $r_3=0.3$ are at $Y=0$, $X=\pm\delta$, $\delta=0.5$. The configuration is defined by $P=S^{\delta+r_3 e^{\imath5\pi/9}}_{\delta+r_3 e^{\imath4\pi/9}}\allowbreak S^{-\delta+r_3 e^{-\imath41\pi/72}}_{-\delta+r_3 e^{-\imath49\pi/72}} \allowbreak S^{e^{-\imath5\pi/11}}_{e^{-\imath\pi/2}} \allowbreak S^{e^{\imath6\pi/11}}_{e^{\imath\pi/2}}\allowbreak B_{0.87}^{0.65\imath}\allowbreak B_{-0.05+0.16\imath}^{0.7\imath}\allowbreak B_{0.05-0.16\imath}^{0.7\imath}\allowbreak B_{0.87e^{\imath\pi/6}}^{-\imath/2}\allowbreak B_{0.87e^{\imath4\pi/3}}^{-\imath/2}\allowbreak B_{0.85e^{\imath5\pi/6}}^{0.7e^{-\imath\pi/12}}\allowbreak B_{0.88e^{\imath9\pi/10}}^{0.7e^{\imath11\pi/12}}\allowbreak B_{0.85e^{\imath\pi/5}}^{0.7e^{-\imath\pi/12}}\allowbreak B_{0.88e^{-\imath\pi/8}}^{0.7e^{\imath11\pi/12}}$ and~\eqref{eq:f_from_P}.}
\end{figure}

In the case of regions with higher connectivity, the magnetic states can be factored in a similar way~\cite{BM15_eng}, using the Schottky-Klein prime function. We will give the expression for states in  canonical circular regions with excised circles, which can be conformally mapped to other regions of the same connectivity. Topological constraints~\cite{BM17}, however, get progressively more complex as connectivity increases (they are not explicitly known yet for connectivity $>2$) and they also mutate when the region is conformally mapped. Luckily, it is sufficient to control only the vortex positions and chiralities in order to build all the necessary states for the present coding scheme. The antivortices then appear ``automatically'' in the correct positions, satisfying the constraints. To this end, let us introduce two functions --- one for vortices inside the face and the other for half-vortices at its boundaries:
\begin{equation}
 \label{eq:multiconnterms}
 \begin{aligned}
 B_{z_0}^{c}(z) & = \frac{(\omega(z,z_0)\omega(1/z,1/z_0))^{-c}}{
 (\omega(z,1/\overline{z_0})\omega(1/z,\overline{z_0}))^{\overline{c}}}, \\
 S^{z_1}_{z_2}(z) & = \frac{\omega(z,z_1)}{\omega(z,z_2)}
 \end{aligned}
\end{equation}
where $\omega(z,\zeta)$ is the Schottky-Klein prime function~\cite{CrowdyMarshall2007}, $z_0$ is the position of the vortex inside the face of the nano-element with the parameter $c$ controlling its chirality (from left-hand to right-hand and also from source to sink type), $z_1$ and $z_2$ are the positions of half-vortices, they must lie on the face's boundary (internal or external). The function $\omega$ depends not only on its arguments, but also on the connectivity and the shape of the nanoelement's face. For a simply-connected region $\omega=\omega_1=z-\zeta$. The function $f$, defining the magnetization distribution (within the assumptions of the model~\cite{M10}) in a multiply-connected region with specified positions of vortices at the boundaries and in the bulk, can be obtained from a product $P$ of any number of functions~\eqref{eq:multiconnterms} by taking a logarithmic derivative
\begin{equation}
\label{eq:f_from_P}
\frac{2\pi}{f} = \frac{\ud \log P}{\ud z}.
\end{equation}
A particular example is shown in Fig.~\ref{fig:triplyconnected}
Note that the central antivortex in the Figure does not code any information and must be present because of the constraint~\cite{BM15_eng,BM17} that the number of antivortices minus number of vortices much be equal to the connectivity of the region minus $2$. The displayed state corresponds to the graph
\begin{equation}
 \vspace{-0.25cm}
 \label{eq:state3c}
 {\tikzmarknode{Bs}{B}\tikzmarknode{Bm}{\overline{B}}\tikzmarknode{Be}{\overline{B}}}\qquad{\tikzmarknode{Cs}{\overline{C}}C\overline{C}\tikzmarknode{Ce}{\overline{C}}\qquad{\tikzmarknode{As}{\overline{A}}A\overline{A}\tikzmarknode{Am}{\overline{A}}A\tikzmarknode{Ae}{\overline{A}}}}.
\end{equation}
\tikz[remember picture, overlay]{
\draw[latex-latex,red] ([yshift=0.1em]As.north) to[bend left] ([yshift=0.1em]Ae.north);
\draw[latex-latex,red] ([yshift=0.3em]Bs.north) to[bend left] ([yshift=0.1em]Be.north);
\draw[latex-latex,red] ([yshift=0.1em]Cs.north) to[bend left] ([yshift=0.1em]Ce.north);
\draw[latex-latex,red] ([yshift=-0.1em]Bm.south) to[bend right=12] ([yshift=-0.1em]Am.south);
\draw[latex-latex,red] ([yshift=-0.1em]Be.east) to ([yshift=-0.1em]Cs.west);
\draw[latex-latex,red] ([yshift=-0.1em]As.west) to ([yshift=-0.1em]Ce.east);
}

It should be noted that the above defined factorizations for the magnetic states fully take into account the exchange interaction, given by the first term in~\eqref{eq:energy}. For any fixed positions of the vortices/antivortices -- every spin in these magnetization vector fields (including those in Figs.~\ref{fig:annulus}, \ref{fig:triplyconnected}) is the solution of the Euler's equation for the exchange energy functional. The magnetostatic interaction is treated only approximately, based on the pole avoidance principle.

Both examples show the magnetization distributions only up to a homotopy (continuous deformation). In reality, when a specific nanoelement's material, size and thickness are taken into account, the vortices and antivortices would deform and shift before they assume their equilibrium positions and shapes. The equilibrium configuration can be computed using the Ritz method from the above trial functions. Or, these functions with suitable choice of initial vortex positions and chiralities can be used to seed a finite element micromagnetic computation.

Provided the boundary half-vortices/half-antivortices remain at their boundaries and vortices do not annihilate with antivortices, there is no way to smoothly deform the magnetization vector field, which would distort the stored graph of symbols. The links correspond to the magnetic domains stretching from one boundary to another. They are also topologically protected because the number of full rotations of the magnetization vector along any contour through the interior of nanoelement's face is preserved with respect to all of its continuous deformations. Note, that the standalone symbols (pairs of half-vortices/half-antivortices bound to the same boundary) do not produce a full rotation of the magnetization vector anywhere inside the face (excluding its boundary), only an incomplete waving.

Let us now discuss the energy barriers.

The vortex/antivortex annihilation barrier is infinite in the continuum model, because of the magnetization vector field singularity, associated with change of the topological index. In reality, due to the discreteness of the material, the barrier for the soliton number change is finite and is determined by interplay of the exchange and the magnetostatic interactions~\cite{PLJU2022}. It is high enough that cross-tie magnetic domain walls -- linear chain crystals of vortices and antivortices can be stable~\cite{HSG1958}.

Half-vortices/half-antivortices can not simply move away from the boundary (or be pushed out of the nanoelement), since this would necessarily produce the side magnetic charges and their magnetostatic energy is strictly positive. The corresponding barrier is finite and high enough to hold the individual side-bound vortices/antivortices, stabilizing head to head domain walls (such as the vortex walls~\cite{Bisig2013VW}) in planar nanorings.

Finally, the most important barrier (making the topological index non-abelian) guards against the merger of two half-vortices/half-antivortices into a complete vortex/antivortex, which is detached from the boundary and moves to the interior of the nanoelement. This process can take place without formation of side magnetic charges. The barrier in this case is created by the interaction energy of the half-vortex/half-antivortex cores (their magnetic poles on the faces of the nanoelement), which have the same polarities and, thus, repel each other. Its height is estimated in the Appendix A. It is also known that the quasi-uniform ``C'' magnetization state in nanodisks with similar energy balance is stable in wide range of disk sizes~\cite{ML08} (which is also confirmed numerically and experimentally at room temperature in a more recent work~\cite{MartinezPerez2020}).

Not only the above-discussed energy barriers exist separately, but recent experiments~\cite{Hempe2007} and simulations~\cite{EL2015} on nanostrips demonstrate the existence and stability at room temperature of similar metastable bound states of vortices and antivortices (including the ones, pinned at the boundary). A simply-connected nanostrip is a trivial limiting case of the present consideration, but the energy barriers, protecting the states in the nanostrip, are essentially the same. These observations are a strong indication that the described states can be realized in practice.

In general, the energy landscape is complex and, because of the long-range magnetostatic forces, depends on the overall shape of the nanoelement. Its precise assessment and optimization can be an interesting problem for numerical modeling. The higher are the barriers, the longer bit strings a particular nanoelement can hold.

Concluding, a coding scheme is introduced, which allows to represent a set of interlinked bit strings as magnetic states of multiply-connected planar nanoelements. Because of the links, multiply-connected nanoelements can hold more information, compared to a set of disconnected simply-connected planar magnetic nanoelements of the same total connectivity. The analytical expressions for the corresponding magnetization distributions in doubly- and general multiply-connected case are developed up to a homotopy. Their application is illustrated for two specific linked bit string sets in doubly- and triply-connected cases. Finally, the energy barriers, protecting the information-bearing states are discussed. Engineering these barriers can be an interesting challenge for future work.

The support of the Russian Science Foundation under the project RSF~21-11-00325 is gratefully acknowledged.

\appendix
\section{vortex split energy barrier}

Let us now estimate the energy barrier, guarding against the split of the magnetic vortex, which is pushed up to the nanoelement's boundary, into two boundary-bound half-vortices.
\begin{figure}
 \begin{center}
 \includegraphics[width=\columnwidth]{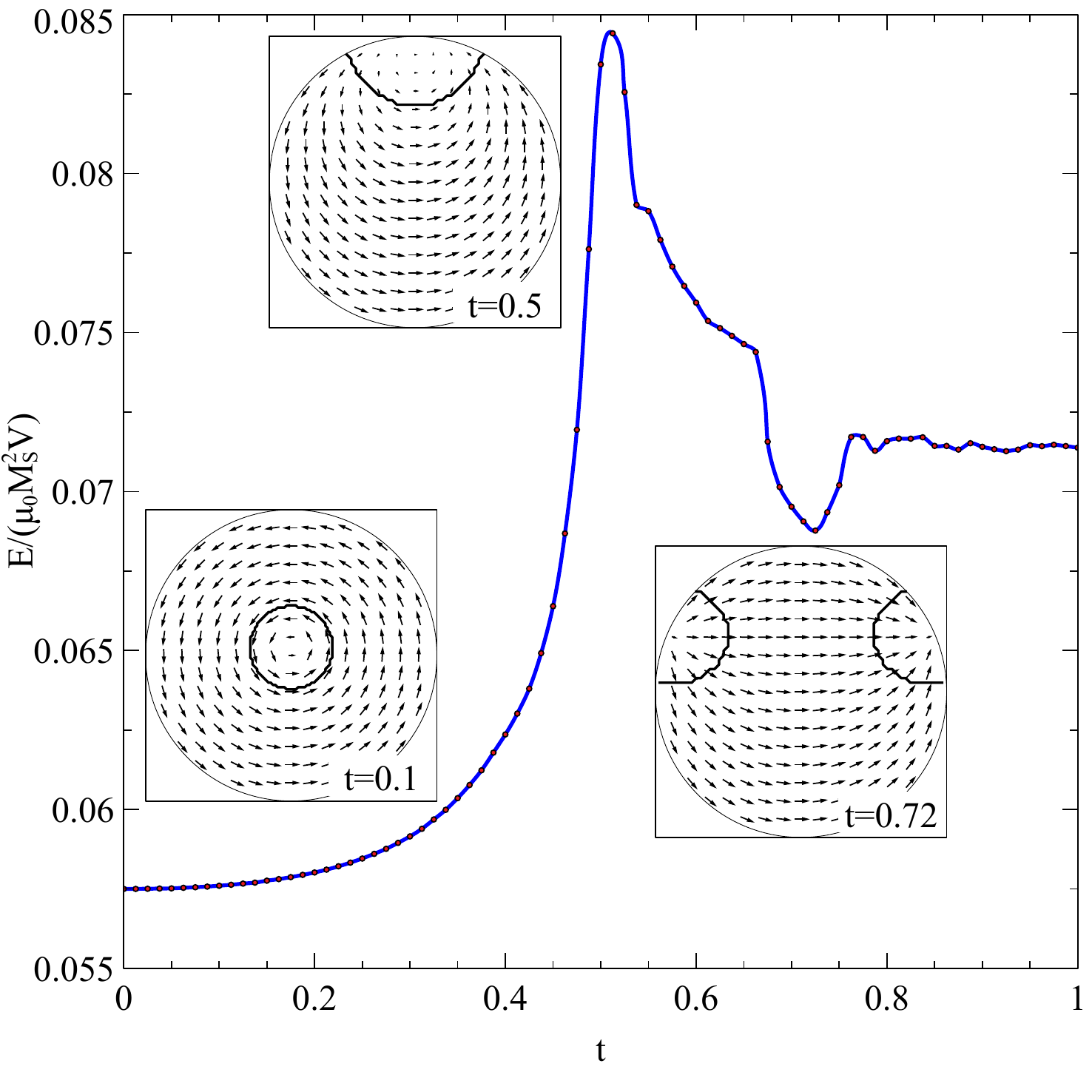}
 \vspace{-1.0cm}
 \end{center}
 \caption{\label{fig:vortexSplit} Normalized total magnetic energy of the magnetic nanodisk with $g=L/R=0.1$ and $\lambda=L/\uLe=0.8$ as a function of vortex displacement/split parameter $t$. Insets show the magnetization distributions ~\eqref{eq:vortexSplit} at chosen values of $t$ with a thick contour inside the particle showing the vortex core boundary.}
\end{figure}
Or, conversely, against the process of merging two half-vortices at the boundary into a complete vortex entering the nanoelement.

%In general, this is a very complex problem with many parameters -- specific particle shape, size, its complete magnetic configuration, material constants. Nevertheless, there is a simple argument for the existence of the barrier for at least some of combinations of these parameters. 
%
%There are three forces at play -- the exchange, the magnetostatic interaction of the face magnetic charges (proportional to the normal magnetization component to the nanoelement's face, which is non-zero inside of the vortex/antivortex cores) and the magnetostatic interaction of the volume magnetic charges (proportional to the divergence of the magnetization, which is too mainly localized around the cores, where the magnetization changes the most).
%
The exchange energy alone does not produce an energy barrier for the vortex at the boundary, it merely tries to push the vortex out of the nanoelement to make its configuration as uniform as possible. The face charges have the same polarity inside of the two boundary-bound half-vortex/half-antivortex cores and thus provide a repulsive force, while the volume charges  have the opposite polarity and produce an attractive force. However, the dependencies of these two opposing forces on the parameters of the problem (most notably on the nanoelement thickness) are different, making it possible to shape the energy landscape by geometry selection, producing and controlling the necessary barrier for some geometries.

For a specific estimate, consider a simplest possible example of a disk-shaped nanoelement (with the radius $R$ and the thickness $L$) in a magnetic configuration, described by a complex function
\begin{equation}
 \label{eq:vortexSplit}
 f(z)=s\left(
   \imath (1 - t) z +
   \frac{t - z^2 \overline{t}}{2}
 \right),
\end{equation}
where $s$ controls the vortex core size (it absorbs the parameter $c_1$ in~\eqref{eq:wpieciwise}, so that we can set $c_1=1$ and $c_2\rightarrow\infty$) and $0\leq t \leq1$ its center displacement as well as the distance between the split half-vortices. This function is essentially the same as in~\cite{M10}, but in a slightly different parametrization to keep the vortex core size roughly the same as parameter $t$ changes. At $t<1/2$ the particle has a vortex inside, at $t=1/2$ this vortex is pushed against the boundary and at $t>1/2$ it is split in two boundary-bound half-vortices. (An example with antivortex splitting is necessarily more complex and is bound to include even more variables, but the energy balance between the volume and the face charges is very similar.)

The total energy~\eqref{eq:energy} of the configuration~\eqref{eq:vortexSplit} in units of $\umuZ \uMssq V$, $V=\pi R^2 L$ is a function of four dimensionless parameters: $t$, $s$, $g=L/R$ and $\lambda=L/\uLe$. Its dependence on $t$, computed by direct numerical integration and minimization (separately for each value of $t$) of the total energy over $s$, for a particular values of $g=0.1$ and $\lambda=0.8$ is shown in Fig.~\ref{fig:vortexSplit}. The full computation is given in a supplemental Wolfram Mathematica notebook~\cite{suppmathfile}.

One can see that there are two stable configurations: one with the magnetic vortex inside of the particle at $t<1/2$ and the other with two halves of the vortex at the boundary at $t>1/2$, separated by a sharp energy barrier. Its height is roughly around $\Delta e \approx 0.017$ in the dimensionless units of the  Fig.~\ref{fig:vortexSplit}. Assuming that the disk is made of a  (permalloy-like) material with the saturation magnetization $\umuZ\uMs=1\mathrm{T}$ and the exchange length $\uLe=5.7\mathrm{nm}$, the height of the barrier is $\Delta e\umuZ\uMssq\pi \uLe^3  \lambda^3/g^2 \approx 4.0\cdot10^{-19}\mathrm{J}$, which is still two orders of magnitude above $k_\mathrm{B} T \approx 4.1\cdot10^{-21}\mathrm{J}$ at room temperature.

%\bibliographystyle{jetpl}
%\bibliography{klm_base}

\end{document}

\maketitle

Современная концепция беговой памяти~\cite{ParkinRacetrack2008} широко используется в скирмионике~\cite{FCS2013} (на ферромагнитных~\cite{TMZTCF2014} и, в последнее время, антиферромагнитных скирмионах~\cite{QLCWF2022}). Информация в ней кодируется как последовательность магнитных доменных границ/скирмионов в нанополосе. Здесь мы обобщим эту идею на многосвязный случай. Возможность последовательного считывания бит при этом теряется, но состояние элемента все же можно оценить по его частотной характеристике, как это продемонстрировано в~\cite{JNFPTSB2012}.

Кодирование является топологическим в том смысле, что записанная информация устойчива к непрерывным деформациям (при выполнении граничных условий) распределения намагниченности. Скирмионная беговая память также топологически защищена, но описанная ниже схема позволяет (для связности $>1$) хранить больше информации.

Известно, что состояния намагниченности в бесконечном двумерном ферромагнетике разбиваются на топологические классы~\cite{BP75}. Распределения одного класса эквивалентны с точностью до непрерывной деформации (гомотопии), а преобразование между распределениями разных классов требует создания сингулярности в векторном поле намагниченности, которая имеет бесконечную энергию
в континуальной модели ферромагнетика. Классы можно пронумеровать целыми числами, а поле намагниченности представить при помощи рациональных функций комплексной переменной~\cite{BP75}. Топологический индекс (номер класса) — общее количество пар вихрь-антивихрь в векторном поле (число скирмионов). В конечных наноэлементах могут присутствовать неполные пары (дополнительный вихрь~\cite{WWBPMW02} или антивихрь~\cite{MEGKRGUFMCSP2010}), из-за чего число скирмионов становится полуцелым.

Топологическая защита делает память более надёжной, но есть и ограничение, что алгебра целочисленного (или полуцелого) топологического индекса является абелевой. Имеет значение только общее количество вихрей/антивихрей внутри элемента. Если он может содержать от $0$ до $n$ скирмионов, то топологически защищено в нём только $\simeq\log_2 n$ бит информации. Обычный способ решения этой проблемы состоит в том, чтобы сделать магнитную среду многосвязной, используя множество (скажем, $N$) отдельных элементов, содержащих (или нет) вихри/скирмионы. Тогда алгебра топологического индекса перестаёт быть абелевой, так как помимо общего числа скирмионов становится важным --- где именно они расположены. Так можно представить около $(1+n)^N$ различных состояний. Рассмотренная здесь альтернатива состоит в том, чтобы использовать один наноэлемент с отверстиями. Он также является многосвязным с неабелевым топологическим индексом. Но, помимо хранения самих бит, в таком наноэлементе можно представить и ссылки между ними, что еще больше расширяет набор доступных топологически защищенных состояний.

В качестве отправной точки возьмём (нормированную на $\umuZ\uMssq$) плотность энергии ферромагнетика в континуальном приближении
\begin{equation}
 \label{eq:energy}
  e=\uLe^2\sum\limits_{i=\uX,\uY,\uZ}|\uGrad\umi(\uvr)|^2 + \uvhD(\uvr,\{\uvm(\uvr)\})\cdot\uvm(\uvr),
\end{equation}
где $\uLe=\sqrt{C/(\umuZ \uMssq)}$ --- обменная длина, $C=2A$ --- обменная жесткость, $\umuZ$ --- проницаемость вакуума, $\uMs$ --- намагниченность насыщения, $\uvm(\uvr)=\uvM(\uvr)/\uMs$ --- нормированный вектор  локальной намагниченности в точке $\uvr=\{\uX,\uY,\uZ\}$, $\uGrad=\{\partial/\partial\uX,\partial/\partial\uY,\partial/\partial\uZ\}$ --- оператор градиента и $\uvhD=\uvHD/\uMs$ --- размагничивающее поле, созданное распределением намагниченности $\uvm(\uvr)$. Рассматриваемые наноэлементы имеют форму (конечного) цилиндра, основание которого может содержать отверстия. Плоскость $\uO\uX$-$\uO\uY$ декартовой системы координат выберем параллельной основаниям цилиндра, а ось $\uO\uZ$ перпендикулярной им. Предположим, что элемент достаточно тонок, чтобы распределение намагниченности можно было считать однородным по его толщине ($\partial\uvm/\partial\uZ=0$). Стороной наноэлемента назовём поверхность, параллельную оси $\uO\uZ$ и проходящую между основаниями цилиндра вдоль их границы. Многосвязные наноэлементы имеют количество сторон, равное связности основания.

Метастабильные распределения намагниченности $\uvm(\uvr)$, минимизирующие объёмный интеграл от~\eqref{eq:energy}, можно приближённо выразить~\cite{M10} в терминах функции $\uw(\uz,\uzc)$ комплексной переменной $\uz=\uX+\imath\uY$, $\uzc=\uX - \imath\uY$, $\imath=\sqrt{-1}$, представляя векторы намагниченности фиксированной длины $|\uvm|=1$ в виде стереографической проекции $\{\umx+\imath\umy,\umz\}=\{2\uw,1-\uw\uwc\}/(1+\uw\uwc)$. Для скирмионов функция $\uw$ является голоморфной ($\partial\uw/\partial\uzc=0$). В более общем случае магнитных вихрей (меронов~\cite{G78}) она выражается через голоморфную функцию $\uf(\uz)$ в виде:
\begin{equation}
\label{eq:wpieciwise}
\uw(\uz,\uzc) = 
\begin{cases} 
      \uf(\uz)/c_1 & |\uf(\uz)| \leq c_1 \\
      \sqrt{\uf(\uz)/\ufc(\uzc)} & c_1 < |\uf(\uz)| < c_2 \\
      \uf(\uz)/c_2 & |\uf(\uz)| \geq c_2 
   \end{cases},
\end{equation}
где вещественные числа $c_1$ и $c_2$ задают размеры ядер солитонов. Центрам вихрей (скирмионов) соответствуют нули $f(z)$, а центрам антивихрей (антискирмионов) её полюса. Чистые  скирмионы реализуются при $c_1=c_2$. В конечном планарном наноэлементе, функция $f(z)$ выражается~\cite{M10} как решение знаменитой задачи Римана-Гильберта о нахождении голоморфной функции без нормальных компонент к границе области (соответствующей абсолютному минимуму энергии магнитных полюсов на сторонах частицы). Эта (однородная) задача имеет множество решений, которые параметризуются скалярами, связанными с положениями центров вихрей и антивихрей. Для получения конкретного равновесного распределения намагниченности в элементе определенного размера (и из определенного материала), необходимо вычислить полную энергию $E=\uiiint e\ud^3r$ и минимизировать её по $c_1$, $c_2$, скалярным параметрам в $f(z)$. Такой подход Ритца гораздо проще исходной вариационной задачи.

Все метастабильные магнитные состояния в рамках этой модели состоят из двух типов магнитных вихрей и антивихрей: 1) целых, расположенных внутри наноэлемента и 2) разделённых на две закреплённые на его границе половинки. Предположим пока, что они никогда не смешиваются и что целый вихрь/антивихрь, созданный внутри наноэлемента, остается внутри, а половинки на границе там всегда и остаются.

Тогда информацию можно представить в виде  последовательности полувихрей и полуантивихрей на границах многосвязного элемента. Поскольку обход вокруг любой границы должен соответствовать полному (или нескольким) обороту вектора намагниченности, на каждой границе всегда имеется чётное число полувихрей и полуантивихрей. Для кодирования произвольных строк удобно представить символ (обозначим его $A$) парой соседних полувихрей на границе, а анти-символ ($\overline{A}$) парой полуантивихрей. Тогда на каждой границе (включая внешнюю) можно расположить строку из символов и антисимволов (количество таких строк равно связности элемента). Кроме того, поскольку каждый полувихрь и полуантивихрь на границе представляет собой начало идущей внутрь наноэлемента доменной стенки, есть две возможности: либо стенка начинается и заканчивается на той же границе, либо на другой границе. При помощи стенок второго типа можно представить ссылки между символами, если они начинаются с полувихрей/анти-вихрей из пары, кодирующей символ на одной границе, и заканчиваются на другой паре, кодирующей символ на другой границе.

Этим способом можно хранить битовые строки со ссылками между индивидуальными битами. Благодаря ссылкам, объем информации, содержащейся в многосвязном наноэлементе, больше, чем во множестве отдельных элементов той же общей связности.

Чтобы пояснить приведенную выше схему, рассмотрим пару примеров. Аналитическая функция $f(z)$ со скалярными параметрами, задающими положения центров магнитных вихрей и антивихрей, может быть выражена через вещественные мероморфные дифференциалы~\cite{B2017-RU} или явно через функции Шоттки-Клейна~\cite{BM15_eng}. При этом нужно учитывать, что в многосвязных наноэлементах существуют дополнительные уравнения связи~\cite{BM17} на положения вихрей и антивихрей, так что не все они независимы.

\begin{figure}
 \begin{center}
 \includegraphics[width=\columnwidth]{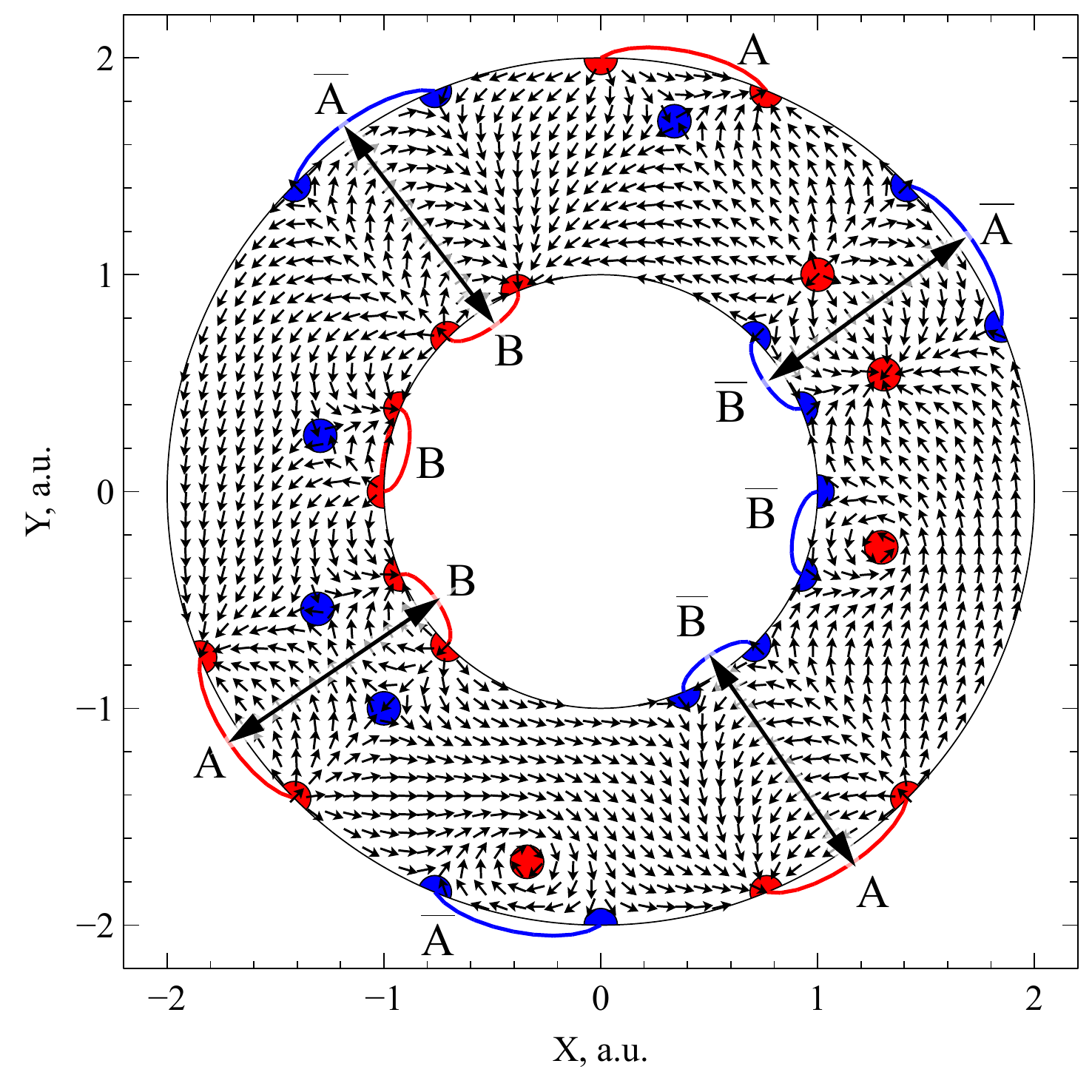}
 \vspace{-1.0cm}
 \end{center}
 \caption{\label{fig:annulus} Кольцо $R=2$ с $N=16$ в конфигурации $f(u)=\imath F_{1,2}^{B}\allowbreak F_{3,4}^{B\overline{A}}\allowbreak F_{5,6}^{A}\allowbreak F_{7,8}^{\overline{A}\overline{B}}\allowbreak F_{9,10}^{\overline{B}}\allowbreak F_{11,12}^{\overline{B}A}\allowbreak F_{13,14}^{\overline{A}}\allowbreak F_{15,16}^{AB}$. Круглыми точками отмечены центры вихрей и антивихрей.}
\end{figure}

Рассмотрим двусвязное планарное концентрическое кольцо внешнего радиуса $R$ и внутреннего $r<R$. Мы не будем сейчас минимизировать полную энергию частицы для определения равновесных значений скалярных параметров модели, а выберем их значения произвольно, что соответствует решению задачи с точностью до гомотопии. В таком случае нет смысла задавать и конкретный пространственный масштаб, поэтому расстояния будем измерять в единицах внутреннего радиуса, положив $r=1$. Рассмотрим конформное отображение $z=Z(u)=\exp(2\pi u/T)$, $T=\pi/\log R$ прямоугольника $0\leq\uRe u \leq1/2$, $-T/2 \leq \uIm u < T/2$ на кольцо $1\leq|z|\leq R$. Внутренняя граница кольца отображается им на $\uRe u=0$, а внешняя на $\uRe u=1/2$. Распределения намагниченности в таком кольце выражаются при помощи периодической с периодом $T$ вдоль мнимой оси голоморфной функции $f(u)$ (на оригинальной комплексной плоскости $f(z)=f(u)Z^\prime(u)$ при $u=Z^{(-1)}(z)$). Задать её можно в терминах эллиптических тета-функций~\cite{B2017-RU} через положения нулей (центров вихрей) и полюсов (центров антивихрей). Уравнения связи~\cite{BM17} легко удовлетворить, если разместить вихри и антивихри с фиксированным шагом по азимутальному углу: в точках $\uIm u = b_n = \imath T/2 - \imath T (n-1)/N$, где $1\leq n\leq N$ и $N$ --- полное количество шагов по азимуту. Точки $u=b_n$ отображаются на внутреннюю границу кольца, а точки $u=a_n=1/2+b_n$ на внешнюю. Введём так же средние точки внутри кольца $c_{n_1 n_2}=1/4+(b_{n_1}+b_{n_2})/2$ и несколько сдвинутых (из эстетических соображений) средних точек $d_{n_1 n_2}=3/20+c_{n_1 n_2}$, $e_{n_1 n_2}=-1/20+c_{n_1 n_2}$. Пару соседствующих половинок вихрей на внешней границе обозначим символом $A$, пару половинок антивихрей на внешней границе символом $\overline{A}$, а такие же пары на внутренней границе символами $B$ и $\overline{B}$ соответственно. Введём 8 функций:
\begin{align*}
F_{n_1 n_2}^{\overline{A}}(u)=&\frac{T_1(u-d_{n_1 n_2})T_1(u+\overline{d_{n_1 n_2}})}{T_1(u-a_{n_1})T_1(u-a_{n_2})}, \displaybreak[0]\\
F_{n_1 n_2}^{\overline{B}}(u)=&\frac{T_1(u-e_{n_1 n_2})T_1(u+\overline{e_{n_1 n_2}})}{T_1(u-b_{n_1})T_1(u-b_{n_2})}, \displaybreak[0]\\
F_{n_1 n_2}^{B\overline{A}}(u)=&-\frac{T_1(u-b_{n_1})T_1(u-\overline{b_{n_2}})}{T_1(u-a_{n_1})T_1(u-a_{n_2})}, \displaybreak[0]\\
F_{n_1 n_2}^{\overline{A}\overline{B}}(u)=& -\left(\frac{T_1(u-c_{n_1n_1})T_1(u+\overline{c_{n_1n_1}})}{T_1(u-a_{n_1})T_1(u-b_{n_1})}\right)\times \\
& \left(\frac{T_1(u-c_{n_2n_2})T_1(u+\overline{c_{n_2n_2}})}{T_1(u-a_{n_2})T_1(u-b_{n_2})}\right),
\end{align*}
$F_{n_1 n_2}^{A}=1/F_{n_1 n_2}^{\overline{A}}$, $F_{n_1 n_2}^{B} = 1/F_{n_1 n_2}^{\overline{B}}$, $F_{n_1 n_2}^{\overline{B}A} = 1/F_{n_1 n_2}^{B\overline{A}}$, $F_{n_1 n_2}^{A B} = 1/F_{n_1 n_2}^{\overline{A}\overline{B}}$,  где $T_1(u)=\theta_1(u,\imath T)$ и $\theta_1(\nu,\tau)$ --- тета функция Ахиезера~\cite{Akhiezer1948-RU} (её можно выразить через тета функцию Якоби как $\theta_1(\nu,\tau)=\imath e^{-\imath\pi(\nu-\tau/4)}\theta(\nu+(1-\tau)/2,\tau)$). Перемножая эти функции, можно представить произвольную конфигурацию битовых строк на границах кольца и ссылки между ними. Функции $F^{A}$,$F^{\overline{A}}$, $F^{B}$,$F^{\overline{B}}$ кодируют несвязанные биты строк, а функции $F^{AB}$, $F^{\overline{A}B}$, $F^{A\overline{B}}$,$F^{\overline{A}\overline{B}}$ кодируют пару бит на разных границах кольца, связанные ссылкой. Пример конфигурации, использующий все определённые выше функции, показан на Рис.~\ref{fig:annulus}.
Он соответствует следующему графу символов
\vspace{0.2cm}
\begin{equation}
 \vspace{-0.2cm}
 \label{eq:state2c}
 {\tikzmarknode{AAs}{\overline{A}}A\tikzmarknode{AAm}{\overline{A}}A\overline{A}\tikzmarknode{AAe}{A}}\qquad
 {\tikzmarknode{BBs}{B}B\tikzmarknode{BBm1}{B}\tikzmarknode{BBm2}{\overline{B}}\overline{B}\tikzmarknode{BBe}{\overline{B}}}.
\end{equation}
\tikz[remember picture, overlay]{
\draw[latex-latex,red] ([yshift=0.1em]AAs.north) to[bend left] ([yshift=0.25em]AAe.north);
\draw[latex-latex,red] ([yshift=0.1em]BBs.north) to[bend left] ([yshift=0.1em]BBe.north);
\draw[latex-latex,red] ([yshift=0.1em]AAm.north) to[bend left=12] ([yshift=0.1em]BBm2.north);
\draw[latex-latex,red] ([yshift=0.1em]AAe.east) to ([yshift=0.1em]BBs.west);
\draw[latex-latex,red] ([yshift=-0.1em]AAs.south) to[bend right=12] ([yshift=-0.1em]BBm1.south);
}
Записанные строки являются циклическими, что обозначено дополнительными ссылками между их началом и концом. Отметить начало строки можно введя дополнительные центры пиннинга для половинок вихрей на границе. Либо можно просто связать начальные биты строк $A$ и $B$ ссылкой. Тогда метки на границе не нужны, но возможностей для кодирования информации ссылками становится меньше.

\begin{figure}
 \begin{center}
 \includegraphics[width=\columnwidth]{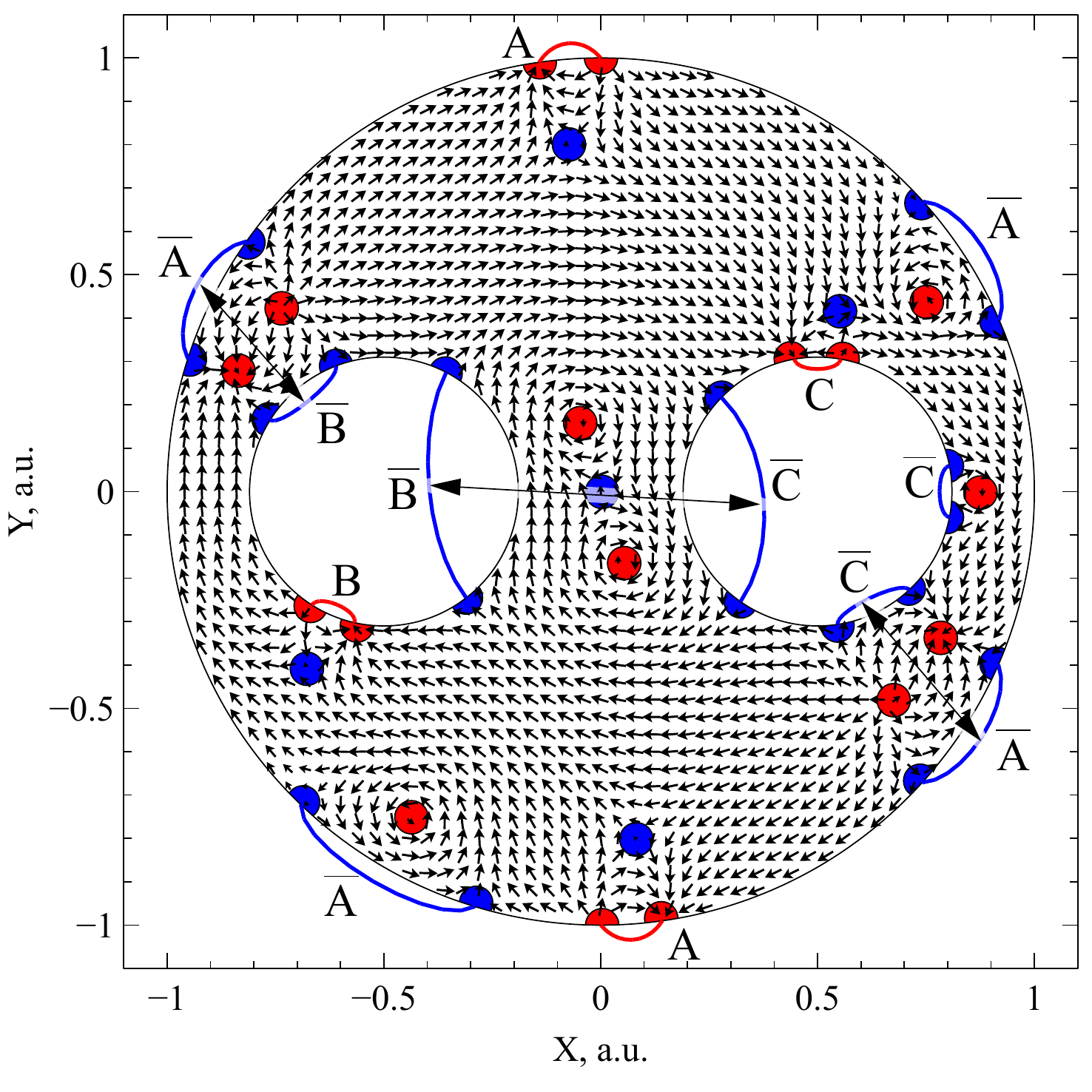}
 \vspace{-1.0cm}
 \end{center}
 \caption{\label{fig:triplyconnected} Трёхсвязный планарный наноэлемент радиуса $R_3=1$ с двумя круговыми отверстиями радиуса $r_3=0.3$ с центрами в $Y=0$, $X=\pm\delta$, $\delta=0.5$. Конфигурация намагниченности определена произведением $P=S^{\delta+r_3 e^{\imath5\pi/9}}_{\delta+r_3 e^{\imath4\pi/9}}\allowbreak S^{-\delta+r_3 e^{-\imath41\pi/72}}_{-\delta+r_3 e^{-\imath49\pi/72}} \allowbreak S^{e^{-\imath5\pi/11}}_{e^{-\imath\pi/2}} \allowbreak S^{e^{\imath6\pi/11}}_{e^{\imath\pi/2}}\allowbreak B_{0.87}^{0.65\imath}\allowbreak B_{-0.05+0.16\imath}^{0.7\imath}\allowbreak B_{0.05-0.16\imath}^{0.7\imath}\allowbreak B_{0.87e^{\imath\pi/6}}^{-\imath/2}\allowbreak B_{0.87e^{\imath4\pi/3}}^{-\imath/2}\allowbreak B_{0.85e^{\imath5\pi/6}}^{0.7e^{-\imath\pi/12}}\allowbreak B_{0.88e^{\imath9\pi/10}}^{0.7e^{\imath11\pi/12}}\allowbreak B_{0.85e^{\imath\pi/5}}^{0.7e^{-\imath\pi/12}}\allowbreak B_{0.88e^{-\imath\pi/8}}^{0.7e^{\imath11\pi/12}}$ и~\eqref{eq:f_from_P}.} 
\end{figure}

В наноэлементах большей связности, магнитные состояния можно факторизовать аналогичным образом~\cite{BM15_eng}, используя функцию Шоттки-Клейна. Рассмотрим здесь канонические круговые области с вырезанными окружностями. Любые другие области могут быть конформно отображены на одну из таких областей той же связности. Каждое отверстие налагает дополнительное ограничение~\cite{BM17} на положения вихрей и антивихрей, которое также видоизменяется при конформном отображении. Явный вид этих ограничений для связности $>2$ еще не известен. К счастью, достаточно управлять только положением и хиральностью вихрей, чтобы построить все необходимые состояния в рамках данной схемы кодирования. Антивихри при этом появляются <<автоматически>> в правильных положениях, удовлетворяющих всем уравнениям связи. Для этого введем две функции --- одну для вихрей внутри наноэлемента, другую для полувихрей на его границах:
\begin{equation}
 \label{eq:multiconnterms}
 \begin{aligned}
 B_{z_0}^{c}(z) & = \frac{(\omega(z,z_0)\omega(1/z,1/z_0))^{-c}}{
 (\omega(z,1/\overline{z_0})\omega(1/z,\overline{z_0}))^{\overline{c}}}, \\
 S^{z_1}_{z_2}(z) & = \frac{\omega(z,z_1)}{\omega(z,z_2)}
 \end{aligned}
\end{equation}
где $\omega(z,\zeta)$ — функция Шоттки-Клейна~\cite{CrowdyMarshall2007}, $z_0$ — положение вихря внутри наноэлемента, параметр $c$ управляет его хиральностью (позволяя вращать все моменты в его окрестности от левостороннего к правостороннему, а также от истока к стоку), $z_1$ и $z_2$ — положения полувихрей на границе (внутренней или внешней). Функция $\omega$ зависит не только от своих аргументов, но также от связности и формы наноэлемента. Для односвязной области $\omega=\omega_1=z-\zeta$. Функцию $f$, определяющую распределение намагниченности (в рамках модели~\cite{M10}) в многосвязной области с заданным положением вихрей на границах и в объеме, можно выразить через логарифмическую производную любого произведения $P$ функций~\eqref{eq:multiconnterms}
\begin{equation}
\label{eq:f_from_P}
\frac{2\pi}{f} = \frac{\ud \log P}{\ud z}.
\end{equation}
Пример такого распределения показан на Рис.~\ref{fig:triplyconnected}
Обратите внимание, что центральный антивихрь на рисунке не кодирует никакой информации и присутствует лишь потому, что количество антивихрей минус количество вихрей равно~\cite{BM15_eng,BM17} связности области минус $2$. Изображенное состояние представляет граф
\begin{equation}
 \vspace{-0.3cm}
 \label{eq:state3c}
 {\tikzmarknode{Bs}{B}\tikzmarknode{Bm}{\overline{B}}\tikzmarknode{Be}{\overline{B}}}\qquad{\tikzmarknode{Cs}{\overline{C}}C\overline{C}\tikzmarknode{Ce}{\overline{C}}\qquad{\tikzmarknode{As}{\overline{A}}A\overline{A}\tikzmarknode{Am}{\overline{A}}A\tikzmarknode{Ae}{\overline{A}}}}.
\end{equation}
\tikz[remember picture, overlay]{
\draw[latex-latex,red] ([yshift=0.1em]As.north) to[bend left] ([yshift=0.1em]Ae.north);
\draw[latex-latex,red] ([yshift=0.3em]Bs.north) to[bend left] ([yshift=0.1em]Be.north);
\draw[latex-latex,red] ([yshift=0.1em]Cs.north) to[bend left] ([yshift=0.1em]Ce.north);
\draw[latex-latex,red] ([yshift=-0.1em]Bm.south) to[bend right=12] ([yshift=-0.1em]Am.south);
\draw[latex-latex,red] ([yshift=-0.1em]Be.east) to ([yshift=-0.1em]Cs.west);
\draw[latex-latex,red] ([yshift=-0.1em]As.west) to ([yshift=-0.1em]Ce.east);
}

Нужно отметить, что определенные выше факторизации для магнитных состояний полностью учитывают обменное взаимодействие, задаваемое первым членом в~\eqref{eq:energy}. При любых фиксированных положениях вихрей/антивихрей - эти поля намагниченности (в том числе и на Рис.~\ref{fig:annulus}, \ref{fig:triplyconnected}) удовлетворяют уравнениям Эйлера для функционала обменной энергии. Магнитостатическое взаимодействие учитывается лишь приближенно за счёт принципа избегания полюсов.

Поскольку минимизация по скалярным параметрам не проводилась, оба примера показывают распределения намагниченности с точностью до гомотопии (непрерывной деформации). В действительности, если принять во внимание материал и размеры наноэлемента, вихри и антивихри будут деформироваться и смещаться, прежде чем займут свои равновесные положения. Эта конфигурация может быть рассчитана по методу Ритца на основе приведенных выше пробных функций. Либо, с подходящим выбором начальных положений вихрей и хиральностей, эти функции можно использовать как затравочную конфигурацию для микромагнитного расчета методом конечных элементов.

Если граничные полувихри/полуантивихри остаются на своих границах, а вихри не аннигилируют с антивихрями --- нет возможности плавно деформировать векторное поле намагниченности так, чтобы исказить сохраненный граф символов. Ссылки между символами соответствуют магнитным доменам, проходящим от одной границы до другой, число которых тоже топологически защищено. То есть количество полных оборотов вектора намагниченности по любому контуру внутри наноэлемента сохраняется по отношению ко всем его непрерывным деформациям. При этом, не связанные ссылками символы (пары полувихрей/полуантивихрей, на одной границе) не дают полного поворота вектора намагниченности вдоль любого контура, кроме точно лежащего на границе частицы.

Обсудим теперь энергетические барьеры.

Барьер аннигиляции вихря с антивихрём в континуальной модели бесконечен из-за сингулярности векторного поля намагниченности, возникающей при изменении топологического индекса. В реальном дискретном материале этот барьер конечен и определяется балансом обменного и магнитостатического взаимодействий~\cite{PLJU2022}. Он достаточно высок, чтобы доменные стенки с перетяжками (линейные кристаллы вихрей и антивихрей) были статически стабильными при комнатной температуре~\cite{HSG1958}.

Полувихри/полуантивихри не могут просто удалиться от границы (или быть вытолкнуты из наноэлемента), так как при этом обязательно возникнут боковые магнитные заряды, а их магнитостатическая энергия строго положительна. Соответствующий барьер конечен и достаточно высок, чтобы удерживать боковые полувихри/полуантивихри во встречных доменных стенках (например, вихревых~\cite{Bisig2013VW}) при комнатной температуре.

Наконец, главный барьер, делающий топологический индекс неабелевым, препятствует слиянию двух полувихрей/полуантивихрей в полный вихрь/антивихрь, входящий внутрь наноэлемента. Такое слияние возможно без образования боковых магнитных зарядов. Барьер в этом случае формируется за счёт взаимодействия торцевых магнитных полюсов в ядрах полувихря/полуантивихря, которые имеют одинаковые полярности и отталкиваются друг от друга. Оценка его высоты дана в Приложении A. Известно также, что квазиоднородное состояние намагниченности типа ``C'' в нанодисках, с подобным балансом энергий, устойчиво в широком диапазоне размеров дисков~\cite{ML08} (что также подтверждается численно и экспериментально при комнатной температуре в работе~\cite{MartinezPerez2020}).

Перечисленные энергетические барьеры существуют не только по-отдельности. Недавние эксперименты~\cite{Hempe2007} и моделирование~\cite{EL2015} на нанополосках показывают существование и устойчивость при комнатной температуре подобных метастабильных связанных состояний вихрей и антивихрей, включая закреплённые на границе половинки. Односвязная нанополоса является тривиальным предельным случаем настоящего рассмотрения, но энергетические барьеры в ней по существу идентичны. Наблюдения~\cite{Hempe2007,EL2015} являются убедительным свидетельством того, что описанные состояния с неабелевым топологическим индексом могут быть реализованы на практике.

В целом, энергетический ландшафт сложен и (из-за дальнодействующих магнитостатических сил) зависит от общей формы наноэлемента. Его точная оценка и оптимизация могут быть интересной задачей для численного моделирования. Чем выше барьеры, тем более длинные цепочки бит может удерживать наноэлемент.

Подводя итог, можно сказать, что в работе вводится схема кодирования, позволяющая представить набор битовых строк со ссылками в виде магнитных состояний многосвязных планарных наноэлементов. Благодаря ссылкам, многосвязные наноэлементы могут содержать больше информации по сравнению с набором изолированных односвязных планарных магнитных наноэлементов той же общей связности. Построены аналитические выражения для соответствующих распределений намагниченности в двусвязном и общем многосвязном случаях с точностью до гомотопии. Их применение проиллюстрировано для двух конкретных наборов битовых строк со ссылками в двух- и трёхсвязном случаях. Обсуждаются энергетические барьеры, защищающие информационные состояния. Оптимизация этих барьеров может стать интересной задачей для дальнейших исследований.

Автор благодарит за поддержку Российский Научный Фонд в рамках проекта 21-11-00325.

\appendix
\section{Барьер деления вихря}

Оценим энергетический барьер, препятствующий расщеплению магнитного вихря, подталкиваемого к границе наноэлемента, на две привязанных к границе половинки.
\begin{figure}
 \begin{center}
 \includegraphics[width=\columnwidth]{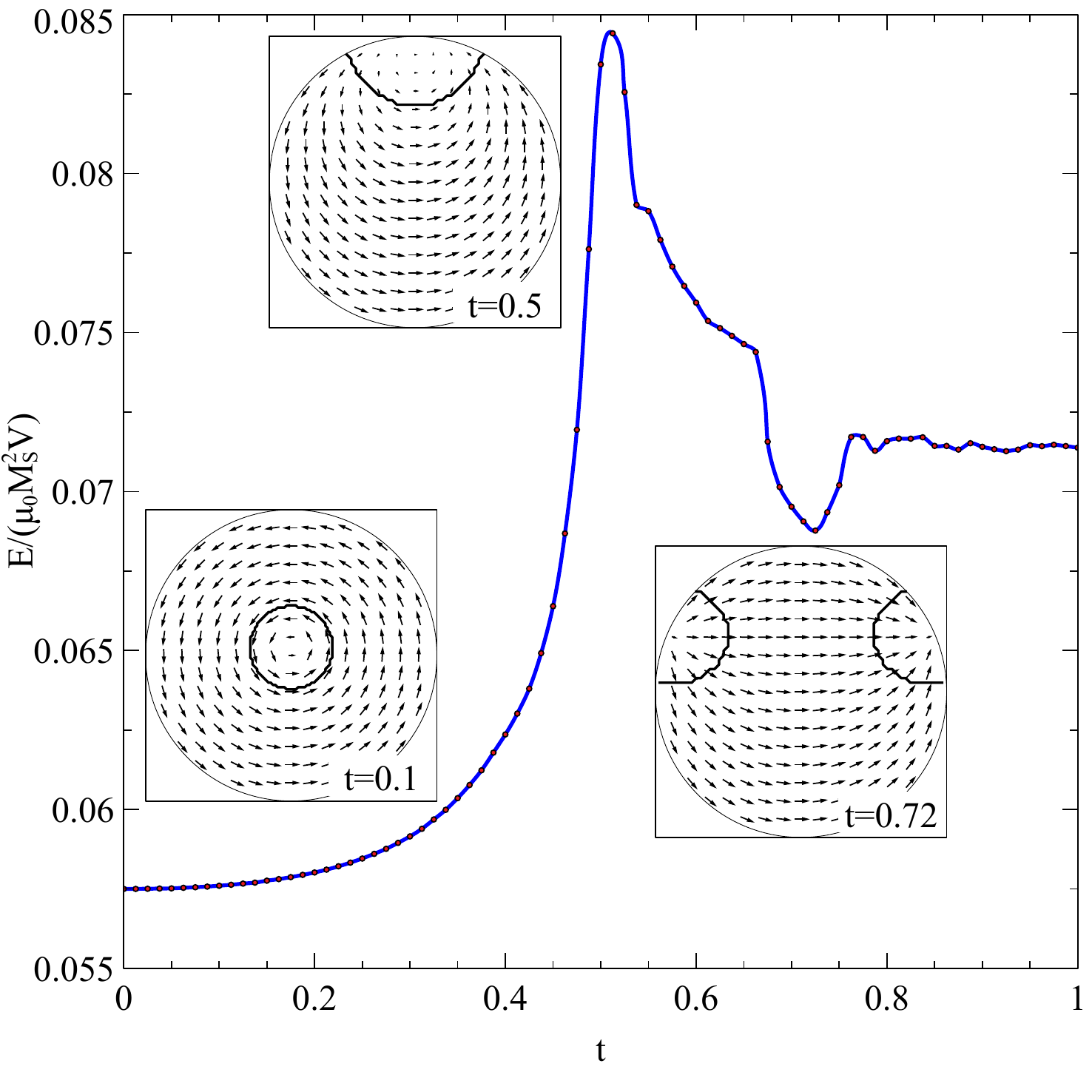}
 \vspace{-1.0cm}
 \end{center}
 \caption{\label{fig:vortexSplit} Нормированная полная энергия магнитного диска с $g=L/R=0.1$ и $\lambda=L/\uLe=0.8$ как функция параметра смещения/разделения вихря $t$. На вставках показаны распределения намагниченности ~\eqref{eq:vortexSplit} при некоторых значениях $t$ (границы ядер вихрей/полувихрей показаны контурами).}
\end{figure}
Или обратно --- процессу объединения двух половинок вихря на границе в полный вихрь с его входом внутрь наноэлемента.

Сама по себе обменная энергия не создает энергетического барьера для вихря на границе, она просто пытается вытолкнуть его из наноэлемента, чтобы сделать конфигурацию максимально однородной. Торцевые заряды внутри двух ядер полувихря/полуантивихря имеют одинаковую полярность и создают силу отталкивания, а объёмные заряды имеют разную полярность и создают силу притяжения. Зависимости этих двух противоборствующих сил от геометрии наноэлемента (в первую очередь от его толщины), однако, различны, что позволяет создавать и контролировать энергетический барьер.

Для конкретной оценки, рассмотрим простейший случай наноэлемента в форме диска радиуса $R$ и толщины $L$ с конфигурацией намагниченности, описываемой комплексной функцией
\begin{equation}
 \label{eq:vortexSplit}
 f(z)=s\left(
   \imath (1 - t) z +
   \frac{t - z^2 \overline{t}}{2}
 \right),
\end{equation}
где $s$ управляет размером ядра вихря (в него входит и параметр $c_1$ из~\eqref{eq:wpieciwise}, так что можно положить $c_1=1$ и $c_2\rightarrow\infty$), а $0\leq t \leq1$ смещением его центра и расстоянием между расщепленными полувихрями. Эта функция по существу такая же, как и в~\cite{M10}, но с несколько иной параметризацией, позволяющей ограничить изменения размера ядра вихря при изменении параметра $t$ во всём интервале. При $t<1/2$ вихрь находится внутри диска, при $t=1/2$ он лежит точно на границе, а при $t>1/2$ расщепляется на два полувихря. (Пример с расщеплением антивихря был-бы неизбежно более сложным из-за необходимости вводить дополнительные переменные, но энергетический баланс между объемными и торцевыми зарядами половинок антивихря аналогичен.)

Полная энергия~\eqref{eq:energy} конфигурации~\eqref{eq:vortexSplit} в единицах $\umuZ \uMssq V$, $V=\pi R^2 L$ является функцией четырех безразмерных параметров : $t$, $s$, $g=L/R$ и $\lambda=L/\uLe$. Её зависимость от $t$, вычисленная прямым численным интегрированием и минимизацией (отдельно для каждого значения $t$) полной энергии по $s$, для $g=0.1$ и $\lambda=0.8$ показана на Рис.~\ref{fig:vortexSplit}. Полное вычисление приведено в дополнительном материале к статье в виде программы для Wolfram Mathematica~\cite{suppmathfile}.

Видно, что существуют две устойчивые конфигурации, разделенные энергетическим барьером: одна с магнитным вихрем внутри частицы при $t<1/2$, другая с двумя полувихрями на границе при $t>1/2$. Высота барьера составляет примерно $\Delta e \approx 0.017$ в безразмерных единицах Рис.~\ref{fig:vortexSplit}. Если предположить, что диск изготовлен из материала подобного пермаллою с намагниченностью насыщения $\umuZ\uMs=1\textnormal{Т}$ и обменной длиной $\uLe=5.7\textnormal{нм}$, высота барьера равна $\Delta e \umuZ \uMssq \pi \uLe^3 \lambda^3/g^2 \approx 4.0\cdot10^{-19}\textnormal{Дж}$, что на два порядка выше $ k_\mathrm{B} T \approx 4.1\cdot10^{-21}\textnormal{Дж}$ при комнатной температуре.

%\bibliographystyle{jetpl}
%\bibliography{klm_base}